\documentstyle[12pt,prd,aps,psfig,floats,preprint]{revtex}

\begin{document}
\title{Using the Magnitude-Squared Coherence for Determining Order-Chaos
Transition in a System Governed by Logistic Equation Dynamics}

\author{Carlos R. Fadragas\thanks{fadragas@mfc.uclv.edu.cu}}

\address{Physics Department, Faculty of Mathematics, Physics and
Computer Science,\\Central University of Las Villas, Santa Clara,
Cuba}

\author{Rub\'en Orozco Morales\thanks{rorozco@fie.uclv.edu.cu}}

\address{Center for Studying Electronics and Information Technologies,\\
Faculty of Electrical Engineering, Central University of Las
Villas, Santa Clara, Cuba}

\date{\today}
\maketitle \draft

\begin{abstract}
This paper is devoted to show the results obtained by using the
magnitude-squared coherence for determining order-chaos transition
in a system described by the logistic equation dynamics. For
determining the power spectral density of a chaotic
finite-duration discrete-time sequence the Welch average
periodogram method was used. This method has the advantage that
can be applied to any stationary signal by using the discrete
Fourier transform (DFT) representation of a discrete-time series
which allows an effective computation via fast Fourier transform
(FFT) algorithm, and that can be applied to a discrete-time series
shorter than that required by nonlinear dynamical analysis
methods. The estimate of the Inverse Average Magnitude-Squared
Coherence Index (IAMSCI) for each discrete-time series in the set
obtained from the logistic mapping was calculated. The control
parameter, {\bf r}, ranges in the interval [2.8,4] for producing
the discrete-time series set. When the condition $r \geq 3.57$ is
satisfied, each discrete-time series exhibited a positive value
for IAMSCI estimate, indicating a high level of coherence loss of
the signal and corresponding to a chaotic behavior of the
dynamical system. Its effectiveness was demonstrated by comparing
the results with those obtained by calculating the largest
Lyapounov exponent of the time series set obtained from the
logistic equation.
\end{abstract}

\newpage
\section{Introduction}
Nonlinear dynamical methods for analyzing a discrete-time series
have been widely used in several fields of scientific research.
For characterizing a scalar discrete-time series metric tools of
nonlinear dynamical methods, such as dimensions, exponents, and
entropies are calculated. Applying these tools to embed a
discrete-time series in a phase space to obtain a phase-space
portrait of the attractor of the dynamical system is required. In
order to make the discrete-time series embedding defining the two
embedding parameters which are the embedding dimension, $D_e$, and
the time delay, $\tau$, is required. Several methods have been
proposed for determining adequate values of these
parameters\cite{Bleek,Takens,Grassberger,Farmer,Eckmann,Kennel,Fojt}.
When the dimensionality, $m$, of the dynamical system is known, it
is directly considered that $D_e=m$, but it is not the case when
we only know a scalar discrete-time series obtained by measuring
one of the accessible variables of the state space of a dynamical
system. The embedding dimension is usually estimate in accordance
with Takens theorem, i.e., $D_e>2*D_2$, where $D_2$ is the
correlation dimension of the attractor in the reconstructed phase
space and it is required to be also estimate before $D_e$ be
known. After determining the two embedding parameters, $D_e$ and
$\tau$, one proceeds to calculate estimates of the metric tools.
Two important requirements must be considered at this point. One
of these is related to the stationarity of the discrete-time
series and the other one is related to the length of the data
point sequence. Nonlinear dynamical methods applying requires that
a discrete-time series to be analyzed be a stationary free-noise
sequence during the observation time. For calculating the
correlation dimension, $D_2$, and the largest Lyapounov exponent,
$\lambda_1$, Eckmann and Ruelle\cite{Eckmann} have suggested a
certain minimal value for the length of the experimental data
points sequence in order for the results obtained by applying
nonlinear dynamical analysis be reliable. It is recognized by
several authors that the application of nonlinear dynamical
methods implies a high computational cost. In this sense,
therefore, any method that allows to obtain any information about
the behavior of a dynamical system and that can operate over a
relatively short discrete-time series, acquires an important
practical value.\\

For complementing those methods of nonlinear dynamical analysis we
shall now try to demonstrate that introducing the concept of
magnitude-squared coherence is possible to quantifying the
order-chaos transition in a dynamical system described by logistic
mapping dynamics. This method is based on the discrete Fourier
transform representation of a finite-duration discrete-time series
and has the advantage that can be applied to a relatively shorter
discrete-time series than that required by nonlinear dynamical
methods. The magnitude-squared coherence has been used by several
authors for measuring constancy of phase between two or more
signals at one
frequency\cite{Thakor,Suresh,WolfE,Eddins,Sih,Liberati}. Besides,
according to literature the transition from a regular to a chaotic
behavior has been studied in many dynamical systems, including the
biological type. One of these is a fluid flow, where by
calculating nonlinear dynamical parameters the order-chaos
transition can be determined.\\

Van den Bleek and Schouten \cite{Bleek} determined the order-chaos
transition in a fluidized bed by calculating both correlation
dimension, $D_{2}$, and Kolmogorov entropy, $K$, as a function of
the Reynolds number and the superficial gas velocity, indicating
two different regimes, order and chaos, and the intermediate
region. In a previous work \cite{Fadragas} the DFT representation
for qualitatively describing the order-chaos transition in a
dynamical system is used. Now we want to make the description of
order-chaos transition in a dynamical system governed by the
logistic equation dynamics by applying the Welch average
periodogram method for calculating the power spectral density and
from this, to introduce an inverse average magnitude-squared
coherence index for indicating quantitatively order-chaos
transition in the dynamical system indicated.\\

The application of magnitude-squared coherence (MSC) method is
reported by several authors. NV Thakor et al\cite{Thakor}
presented the power spectral density analysis of electrocardiogram
(ECG) waveforms as well as isolated QRS complexes and episodes of
noise and artifacts. The power spectral analysis showed that the
QRS complex could be separated from other interfering signals. A
bandpass filter that maximizes the signal-to-noise ratio would be
of use in many monitoring instruments. They calculated the
coherence function and, from that, the signal-to-noise ratio. Upon
carrying out this analysis on experimentally obtained ECG data,
they observed that a bandpass filter with a center frequency of 17
Hz and a Q factor of 5 yields the best signal-to-noise ratio. They
estimated the coherence function as given by

\begin{eqnarray}
C_{xy}^2=\frac{\vert G_{xy}(f).G_{xy}(f)^*\vert^2}{G_{xx}(f).G_{yy}(f)}
\end{eqnarray}

where $G_{xx}(f)$ is the estimates of the auto-power spectra of
the QRS complex, $G_{yy}(f)$ is the complete ECG cycle, and
$G_{xy}(f)$ is the cross power spectrum estimate. For calculating
the power spectra they applied the Fourier transform of the
windowing signal. To improve the resolution of the FFT algorithm
and the power spectra, they filled the original data sets with
zeros and formed new 1024 point data sets. S Narayanaswamy et
al\cite{Suresh} developed three signal processing tools to
identify three postulated mechanisms of isolated or single
premature ventricular contraction (PVC) generation. Two hours of
continuous ECG recording were digitally obtained from several
patients with frequent PVC. They applied the magnitude-squared
coherence (MSC) spectrum between the sinus intervals $x[n]$ and
PVC intervals $y[n]$ as given by

\begin{eqnarray}
MSC(f)=\frac{\vert
S_{x}(f).S_{y}(f)^*\vert}{\sqrt{S_{xx}(f)}.\sqrt{G_{yy}(f)}}
\end{eqnarray}

where $S_{x}(f)$ and $S_{y}(f)$ are the Fourier spectra of $x[n]$
and $y[n]$ respectively. $S_{y}(f)^*$ is the complex conjugate of
$S_{y}(f)$, and $S_{xx}(f)$ and $S_{yy}(f)$ are the auto-power
spectra of $x[n]$ and $y[n]$ respectively. The power spectral
density of the PVC interval series is defined as

\begin{eqnarray}
P(k)=\vert X(k)\vert^2
\end{eqnarray}

where $X(k)$ is the Fourier transform of $x[n]$. In this work,
details of the used method are not showed, but they used both the
power spectral density of the PVC signal and the MSC estimates of
the PVC and sinus intervals.\\

EG Lovett y KM Ropela\cite{Lovett} showed that magnitude-squared
coherence (MSC) of intracardiac electrogram may be used to
quantify what may be termed organization. Unfortunately, MSC
computations are not meaningful from the surface due to the high
correlation among measured body surface potentials. They showed
that average magnitude-squared auto-bi-coherence (AMSABC) provides
an organizational measure of cardiac rhythms from a single surface
lead. They showed that AMSABC also discriminates three ventricular
tachyarrthythmia. In this work, rather than measuring consistency
of phase between two signals at one frequency (as in MSC),
measuring consistency between two frequencies in one signal (as in
MSABC) were considered. Haris J. Sih et al\cite{Sih} applied the
MSC function for analyzing the spatial organization of epicardial
mapping in the frequency domain. They defined the
magnitude-squared coherence (MSC) as a frequency domain measure of
the phase consistency between two signals, $x[n]$ y $y[n]$, as
given by

\begin{eqnarray}
MSC_{xy}(f)=\frac{\vert S_{xy}(f)\vert^2}{S_{xx}(f).S_{yy}(f)}
\end{eqnarray}

where $S_{xy}(f)$ is the cross-power spectrum, and $S_{xx}(f)$ and
$S_{yy}(f)$ are the respective auto-power spectra, MSC is
dimensionless and can vary from 0 to 1. A MSC value of 1 at some
frequency would indicate a linear relationship or perfect
coherence between the two signals at that frequency, while a value
of 0 would indicate no relationship or perfect incoherence between
the two signals at that frequency. MSC is sensitive to noise and
interference, whereas uncorrelated noise in either of the two
signals decreases MSC, and correlated interference increases MSC.
While the relationship between MSC and correlated interference is
difficult to characterize, the influence of noise on MSC can be
derived for two real signals, $u(t)$ and $v(t)$, equal to ideal
signals, $x(t)$ and $y(t)$, plus random, independent, zero-mean
noise processes, $n_{1}(t)$ and $n_{2}(t)$. It is easy to show
that

\begin{eqnarray}
MSC_{uv}=\frac{MSC{xy}(f)}{1+\frac{1}{SNR_{x}(f)}+\frac{1}
{SNR_{y}(f)}+\frac{1}{SNR_{x}(f)SNR_{y}(f)}}
\end{eqnarray}

where $MSC_{uv}$ is the magnitude-squared coherence between $u(t)$
and $v(t)$, $MSC_{xy}$ is the magnitude-squared coherence between
$x(t)$ and $y(t)$, $SNR_{x}(f)$ is the frequency-dependent
signal-to-noise ratio between $x(t)$ and $n_{1}(t)$, and
$SNR_{y}(f)$ is the frequency-dependent signal-to-noise ratio
between $y(t)$ and $n_{2}(t)$. As expected, as the signal-to-noise
ratios increase, $MSC_{uv}$ approaches to $MSC_{xy}$. $MSC$ can be
estimated by replacing $S_{xy}(f)$, $S_{xy}(f)$, and $S_{xy}(f)$
with respective estimates by using the Carter method. This method
uses average Fast Fourier Transform of weighted overlapped
segments of equal length from the two signals to form the spectral
estimates. That is, $MSC$ is estimated as given by

\begin{eqnarray}
MSC_{xy}(f)=\frac{\vert\sum_{i=1}^N X_{i}(f)Y_{i}^*(f)\vert^2}
{\vert\sum_{i=1}^N X_{i}(f)X_{i}^*(f)\sum_{i=1}^N Y_{i}(f)Y_{i}^*(f)\vert}
\end{eqnarray}

where $X_{i}(f)$ and $Y_{i}(f)$ are the FFT of the $i-th$ weighted
segments from $x(t)$ and $y(t)$, respectively, $N$ is the total
number of segments, and $(*)$ denoted the complex conjugate. From
this method of estimation, one can interpret $MSC$ as a measure of
the phase consistency between the two signals. At a single
frequency $f_{1}$, the cross-power term in the numerator for one
segment can be interpreted as a vector with amplitude $\vert
X_{i}(f_{1})\vert\vert Y_{i}(f_{1})\vert$ and phase $\angle
X_{i}(f_{1})-\angle Y_{i}(f_{1})$. Therefore, the sum of the
cross-power terms in the numerator of the $MSC$ estimate is
equivalent to the vector sum of the cross-power vectors for each
segment. If $\angle X_{i}(f_{1})-\angle Y_{i}(f_{1})=\angle
X_{j}(f_{1})-\angle Y_{j}(f_{1})$ for all $i$ and $j$, that is, if
a constant phase relationship from one segment to next exists
between $x(t)$ and $y(t)$ at $f_{1}$, the $N$ vector terms are all
aligned and the vector sum is at a maximum, as is the $MSC$
estimate. If the terms are not constant from segment to segment,
the $N$ vector terms do not align, the vector sum in the numerator
is not at a maximum, and the $MSC$ estimate must be less than
unity at that frequency. As with all methods of spectrum
estimation, the $MSC$ estimate has statistical bias and variance,
which have been explicitly calculated for the case of Gaussian
stationary signals. Another form of bias, known as a bias due to
misalignment, has also been shown to exist for this estimator. As
an example, for two signals, $x(t)$ and $y(t)=b*x(t-D)+n(t)$,
where $n(t)$ is uncorrelated noise, it has been demonstrated that
the estimate of the true $MSC$ spectrum will be degraded by a
factor of $(1-\vert D\vert/T)^2$ where $T$ is the time duration of
one windowed segment of data. For example, if $T$ were
approximately $0.25 s$, a relative delay of $0.1 s$ between the
two signals could decrease the $MSC$ estimate to one-third the
true value. It is noted that these results are derived for two
linearly related signals, and the effects of this bias due to
misalignment on two nonlinearly related signals are yet to be
determined. With various types of mapping data, as those of
cardiology, of neurology, of geophysics, etc, multiple signals are
recorded simultaneously, and often, it is necessary to preserve
the physical location of these recordings relative to the
underlying activity. Under this conditions, $n$ signals,
$x_{1}(t), x_2(t),\cdot,x_n(t)$, are arranged in a $1-$, $2-$, or
$D-$ array, or map. By choosing some meaningful reference signal,
$X_{r}(t)$, which can be one of the $n$ signals in the map, a
corresponding map of $MSC$ spectra can be formed. That is, what we
have is a $n-MSC$ spectra set: $MSC_{x_{1}x_{r}}(f)$,
$MSC_{x_{2}x_{r}}(f)$,$\cdots$, $MSC_{x_{n}x_{r}}(f)$, arranged in
the same $1-$, $2-$, or $D-$ map. In the case where the reference
is one of the $n$ signals in the map, the $r-th$ $MSC$ spectrum,
$MSC_{x_{r}x_{r}}(f)$, is unity for all frequencies. It is noted
that the map of $MSC$ spectra may indicate how the $n$ signals
relate to the reference, but they do not imply any relationship
between one signal of the $n$ signals with any other of the $n$
signals. Different maps of $MSC$ spectra can be generated using
different references. A logical and concise method for displaying
these $MSC$ spectra is problematic as pointed out by many authors.
To simplify the data, one can reduce each $MSC$ spectrum to a
single number, for example, by averaging $MSC$ over frequency.\\

Finally, we refer to the work done by D Liberati et
al\cite{Liberati}, where the joint use of total and partial
coherence between pairs of $EEGs$ simultaneously recorded in a
standard set is shown to enhance what is caused direct correlation
between cortical subsystems and what is instead related to the
spread of the electromagnetic field. A multi-variable $AR$
approach is employed in the computation, giving results even for a
very short time window, thus allowing coherence to be investigated
at the main cortical latencies of evoked potentials. In
particular, when a combined visual and somatosensory stimulation
is applied, cortical interactions are captured in the frequency
domain. In this work, the total squared coherence $k$, as a
function of the frequency $f$ is defined, for every couple $j,h$
of the channels, as the squared cross-spectrum $P_{jh}$,
normalized to the product of the norm of each auto-spectrum
$P_{jj}$ and $P_{hh}$ as given by

\begin{eqnarray}
k_{jh}(f)=\frac{\vert P_{jh}(f)\vert^2}{\vert P_{jj}(f)\vert.\vert P_{hh}(f)\vert}
\end{eqnarray}

where $k_{jh}$ is a real function, ranging from 0 to 1, indicating
the amount of correlation between the signals $j$ and $h$ as a
function of the frequency $f$. In this work we pretend to
demonstrate that it is possible to apply the average Welch
periodogram for estimating the power spectral density of a chaotic
signal and to calculate the magnitude-squared coherence for
determining quantitatively the order-chaos transition in a
dynamical system described by the logistic mapping dynamics. The
fundamental advantage of this method upon the others is that it
can operate on a relatively shorter discrete-time series. By
estimating the largest Lyapounov exponent was able to evaluate
those results obtained here applying an inverse average
magnitude-squared coherence index, defined by averaging $MSC$
estimate over the frequency.

\section{Methods and Materials}

\subsection{Obtaining the discrete-time series set}
The logistic equation is an one-dimensional mapping of the real
axis in the interval (0,1) to itself, that is a prototype of a
nonlinear dynamical system widely used\cite{Cohen} and can be
formulated as

\begin{eqnarray}
x[n+1]=rx[n](1-x[n]),
\end{eqnarray}

and it can also be formulated as

\begin{eqnarray}
x[n+1]=r^n(x[1]-\sum_{k=0}^nr^{-k+1}x[k+1]);
\end{eqnarray}

and with the initial condition $x[1]$ specified, one can obtained
a time sequence of a given length by a recursive action. The
character of the behavior of a data sequence obtained with the
iterative process can be modified by conveniently selecting a
value of the control parameter $r$. In this work a value of the
initial condition $x[1]=0.65$ was taken, and the interval $r \in
[2.8,4.0]$ was selected. A set of evenly spaced values of $r$,
with a step $\delta=0.01$, was taken to produce a family of 120
discrete-time series, each containing $N=2^{10}$ samples. The
building up of the histogram for each time series allows to have
immediately a simple statistical characterization of this time
series. In order to make easier further computations data points
were organized into a rectangular matrix containing 120 columns,
each of which is a time series with the length $N=1024$ data
points.

\subsection{Determining the largest Lyapounov exponent $\lambda_1$}

Lyapounov exponents spectrum deals with average exponential
divergence or convergence of two neighbor orbits in the phase
space. It is common to order the spectrum from largest to shortest
exponents: $\lambda_1,\lambda_2,\lambda_3,\ldots$. A system
containing one or more positive exponents is to be defined as
chaotic. For calculating the Lyapounov exponents spectrum some
algorithms have been developed, according to the particular
situation, being one of the most used algorithms that reported by
Wolf et al\cite{Wolf}. From the exponents spectrum, the largest
exponent, $\lambda_{1}$, decides the behavior of the dynamical
system. For calculating the largest exponent the algorithm
proposed by Rosenstein et al\cite{Rosenstein} can be referred too.
Details of these algorithm can be found in referred papers. We
determined the largest Lyapounov exponent,$\lambda_1$, for each
time series in the set, using a professional
software\cite{Sprott}. The selection of the time delay, $\tau$,
and the embedding dimension, $D_e$, required for reconstructing
the phase space is part of the problem and it is necessary to
consider some selecting criteria. This selection can be made by
different
methods\cite{Bleek,Takens,Grassberger,Farmer,Eckmann,Kennel,Fojt}.
For selecting $\tau$ we may take into account:(a) optimal filling
of the phase space\cite{Buzug}, (b) by determining the position of
the first local minimum of the data autocorrelation
function\cite{Albano}, (c) by taking the first minimum in the mean
mutual information plot\cite{Fraser}, (d) based on the optimal
tradeoff between redundance and the irrelevance\cite{Rosenstein},
and by analyzing the differential equations of the system if they
are given\cite{Yang}. For selecting $D_e$ it may be indicated: (a)
by applying the Grassberger-Procaccia method, which allows to
obtain simultaneously both correlation dimension and embedding
dimension\cite{Grassberger}, (b) by applying the false nearest
neighbors method\cite{Rosenstein,Blanco}, and (c) by applying
geometric considerations\cite{Kennel}. In this work, a time series
obtained from the logistic equation was analyzed by using the
method (a) for $\tau$ determination and by taking a value $D_e=3$,
which satisfies the Grassberger-Procaccia algoritm condition
$D_e>2D_2$, and Takens theory\cite{Grassberger,Takens}, being
$D_2=0.5$ a value reported by Hoyer et al\cite{Hoyer} for this
time series type. Literature refers\cite{Grassberger,Takens} that
for selecting the embedding dimension as

\begin{eqnarray}
D_e \geq D_2+1,
\end{eqnarray}

it is sufficient to have a good representation for the attractor
in the phase space, reconstructed in delayed coordinates from a
scalar time series observed. Any reconstruction made with a value
of the embedding dimension less than the minimal value will
produce a projection, in that dimension, of the original
attractor, and therefore it will be more difficult to understand
that projection or it will produce false results when an
estimation of a nonlinear dynamical parameter is to be made. It is
important to take a value for embedding dimension close to
$2D_2+1$ in order to avoid the computational cost be excessively
high.

\subsection{Estimating the IAMSCI using the Welch average periodogram}

It is highly meaningful when dealing with a chaotic system keeping
in mind that local sensitivity to a small error is the hallmark of
such a system. Such dynamical system as the one described by the
logistic equation produces a discrete-time non-chaotic series when
the control parameter $r$ value satisfies the condition $r<3.57$.
For such a condition a discrete-time series generated by the
recursive process does not almost change if the initial condition
is modified and there exists phase consistency between the two
discrete-time series at a given frequency. However, when the
control parameter value is greater than 3.57, a discrete-time
series generated by the logistic equation does meaningfully change
if the initial condition is slightly modified and phase
consistency between the two discrete-time series at a given
frequency, in general, must be lost. This effect may be quantified
by applying an inverse average magnitude-squared coherence index.
In order to apply the Welch average periodogram method for
determining the power density spectrum $S_x[k]$ of a discrete-time
series one follows the following steps:\\

(a) Decompose the sequence of N (=1024) data points in M (=9)
segments, each one with the same length L (=256), and which may be
overlapped P samples. The
usual value $P=0.625*L$ was chosen.\\

(b) Calculate the DFT representation by FFT algorithm for each
segment as given by $X_m[k]=fft(x_m[n])$, where $m=1,2,\ldots,M$,
and $n=0,1,2,\ldots,L-1$, and fft represents a function in MatLab
for calculating the DFT of a given signal. For each segment a
L-FFT was calculated,
being L the number of points in the sequence $X_m[k]$.\\

(c)Calculate the periodogram for each segment $m$ as given by

\begin{eqnarray}
P_m[k]=\frac{1}{L}\vert X_m[k]\vert^2,\;\;m=1,2,\ldots,M
\end{eqnarray}

(d)Finally, calculate the average periodogram as given by

\begin{eqnarray}
S_x[k]=\frac{1}{M}\sum_{m=1}^M P_m[k]
\end{eqnarray}

where $k=0,1,2,\ldots,L-1$. For a short discrete-time series a
rectangular window is recommended. For determining the
magnitude-squared coherence sequence the procedure described above
is applied to a second signal $y[n]$ obtaining the average
periodogram $S_y[k]$, and the average cross-periodogram
$S_{xy}[k]$ is also calculated. The magnitude-squared coherence
sequence was calculated as given by

\begin{eqnarray}
MSC_{xy}[k]=\frac{S_{xy}[k]*S_{xy}^*[k]}{S_{x}[k]*S_{y}[k]}
\end{eqnarray}

The $MSC_{xy}[k]$ sequence describes the phase consistency between
the two signals at a given frequency $k$. The mean value of the
sequence, $(\langle MSC_{xy}[k]\rangle)$, may be used as an index of
global coherence between the two signals. When dealing with a
chaotic discrete-time series this index must tend to zero, and
when dealing with a deterministic discrete-time series this index
must tend toward one. However, it was found that a better index
may be defined as given by

\begin{eqnarray}
f=10*log_{10}\frac{1}{\langle MSC_{xy}[k]\rangle}.
\end{eqnarray}

It makes easier to compare the results obtained here with those
obtained by applying the largest Lyapounov exponent estimate.

\section{Results}

Figures, from 1 to 4, show a sample of four representative time
series and their corresponding magnitude-squared coherence
spectra. In order to give a better view of each time series it was
only considered 128 data points of each discrete-time series in
its plotting, but it was completely considered when its
corresponding magnitude-squared coherence was calculated. In
either case, the particular values of $r$ parameter were
indicated. Figure 5 depicts results of plotting the inverse
average magnitude-squared coherence $(1/\langle
MSC_{xy}[k]\rangle)$, calculated for each time series in the set,
as a function of the control parameter $r$. And figure 6 shows the
results of plotting the inverse average magnitude-squared
coherence index, in decibels $(f=10*log_{10}(1/\langle
MSC_{xy}[k]\rangle)$, calculated for each time series in the set,
as a function of the control parameter $r$ too. Figure 7 shows
results of plotting the largest Lyapounov exponent estimate,
$\lambda_1$, calculated for each time series in the set, as a
function of the control parameter $r$. This reference serves as a
control for the discussion of the results obtained by applying the
inverse average magnitude-squared coherence index method.

\section{Discussion of the results}

The application of the DFT representation of a chaotic-type signal
gives satisfactory results for determining the order-chaos
transition in a system described by the logistic equation
dynamics\cite{Fadragas}. A critical value for control parameter
$r$ is reported in literature beyond which a sequence produced by
the logistic equation exhibits a chaotic behavior. This threshold
value approaches $r\cong 3.57$. On the other hand, figures from 1
to 4 show the time domain representation of some of the time
series and their corresponding magnitude-squared coherence
spectrum. Comparing figures 1 ($r=2.91$) with 2 ($r=3.64$), 3
($r=3.75$), and 4 ($r=3.84$), it can be deduced the bifurcation
effect and the transition to chaos. Note that in figures 2, 3 and
4 the control parameter satisfies the condition $r>3.57$, and it
corresponds to a value for a positive largest Lyapounov exponent,
as it can be estimated from figure 7. It may be observed in those
cases that the magnitude-squared coherence spectrum exhibits
values far away from one for several values of the frequency in
the $MSC$ spectrum. It makes the average magnitude-squared
coherence to be less than one, corresponding to a situation for
which the discrete-time series becomes incoherent and indicating in
this case a chaotic behavior of the dynamical system. This
situation can be observed in figures 5 and 6 where either the
quantity $(1/\langle MSC_{xy}[k]\rangle)$ as the inverse
average $MSC$ index, $(f=10*log_{10}(1/\langle
MSC_{xy}[k]\rangle)$, are greater than $1$ and $0$, respectively,
for $r>3.57$. Applying the inverse average magnitude-squared
coherence index (IAMSCI) to a discrete-time series in a family
derived from an observable in a dynamical system is recommended
for quantitative detecting the order-chaos transition, and this
can be added to the metric tools of the nonlinear dynamical
analysis for complementing. Any method that allows to evaluate the
behavior of a dynamical system, and that can be applied to a
relatively shorter discrete-time series, must be taken into
account as an important practical method.

\section{Conclusions}

The application of the inverse average magnitude-squared coherence
index (IAMSCI), $f$, to a discrete-time series obtained from the
logistic equation can be done on a chaotic discrete-time series.
This discrete-time series has the property that exhibits a high
level of loss of coherence, and it can be detected by applying the
IAMSCI method. This method does not exert a strong requirement on
the length of the experimental data sequence. If one deals with a
nonlinear dynamical system which can modify its behavior between
order and chaos, a time series produced by that system reflects
this change, and the IAMSCI method applying to the discrete-time
series allow to determine this change. Any method that permits
some quantitative evaluation about the behavior of a dynamical
system and that can work over a relatively short time series
acquires a practical importance. Besides, this method does not
require the phase space reconstruction of the attractor.

\section*{Acknowledgement}

We acknowledge Pedro A. S\'anchez Fern\'andez from the Department
of Foreign Languages for the revision of the English version, and
Yoelsy Leiva for the edition. We also thank the Ministry of Higher
Education of Cuba for partially financial support of the research.

\newpage
\section{Figures}

\newpage
\begin{figure}[tbh!]
\centerline{\psfig{figure=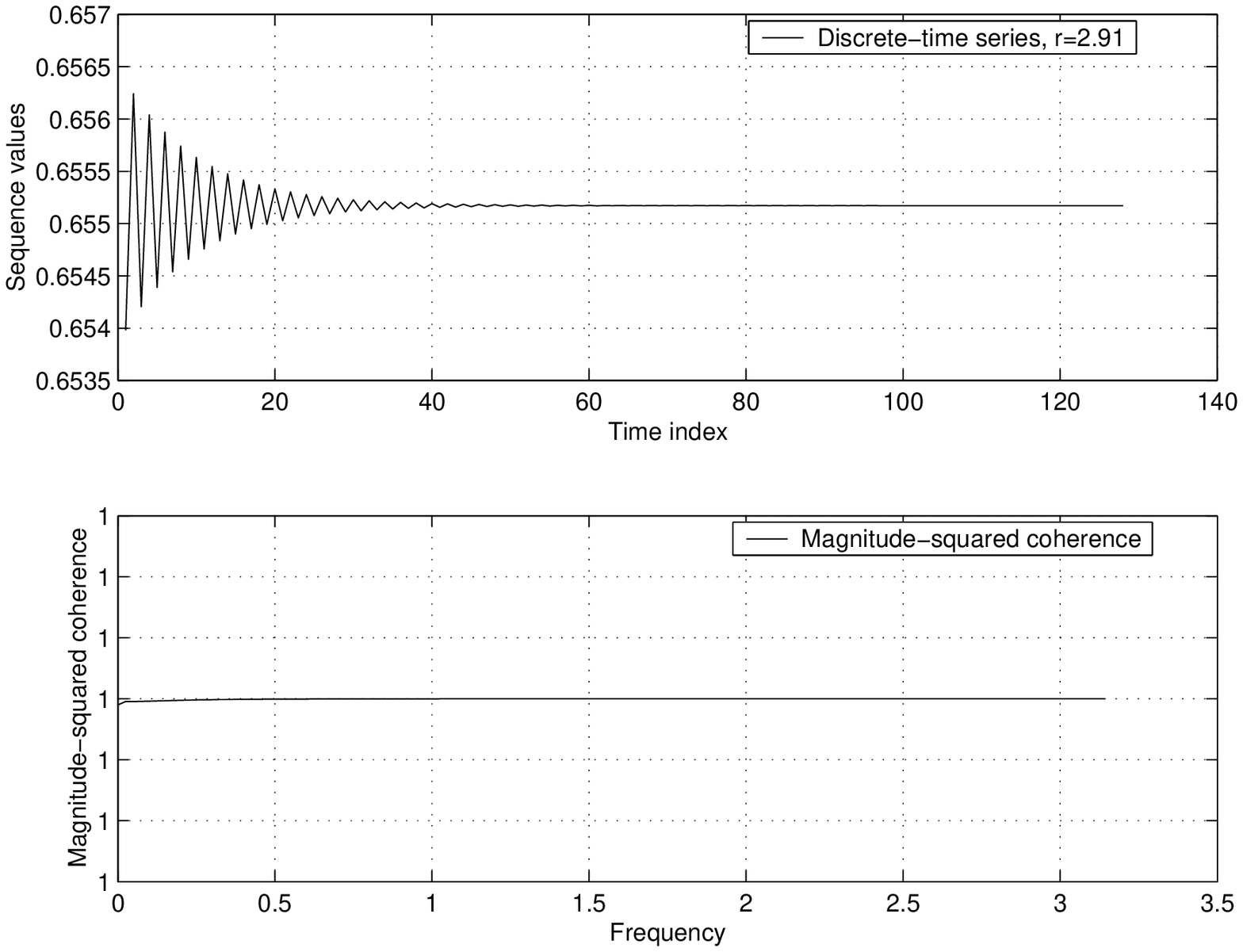,width=0.5\textwidth,angle=0}}
\bigskip
\caption{Discrete-time series and its magnitude-squared coherence,
for r=2.91} \label{fig1}
\end{figure}

\begin{figure}[tbh!]
\centerline{\psfig{figure=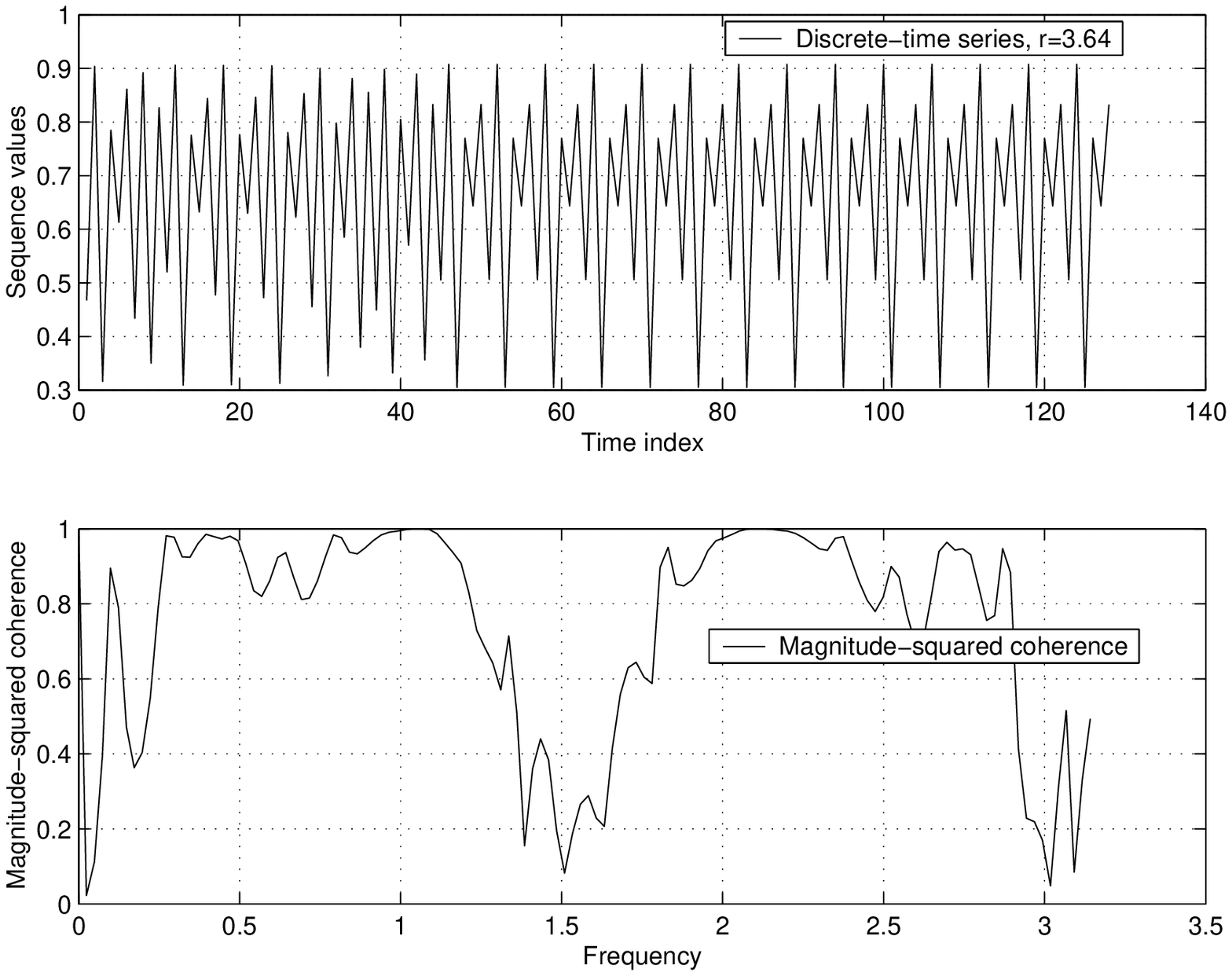,width=0.5\textwidth,angle=0}}
\bigskip
\caption{Discrete-time series and its magnitude-squared coherence,
for r=3.64} \label{fig2}
\end{figure}

\begin{figure}[tbh!]
\centerline{\psfig{figure=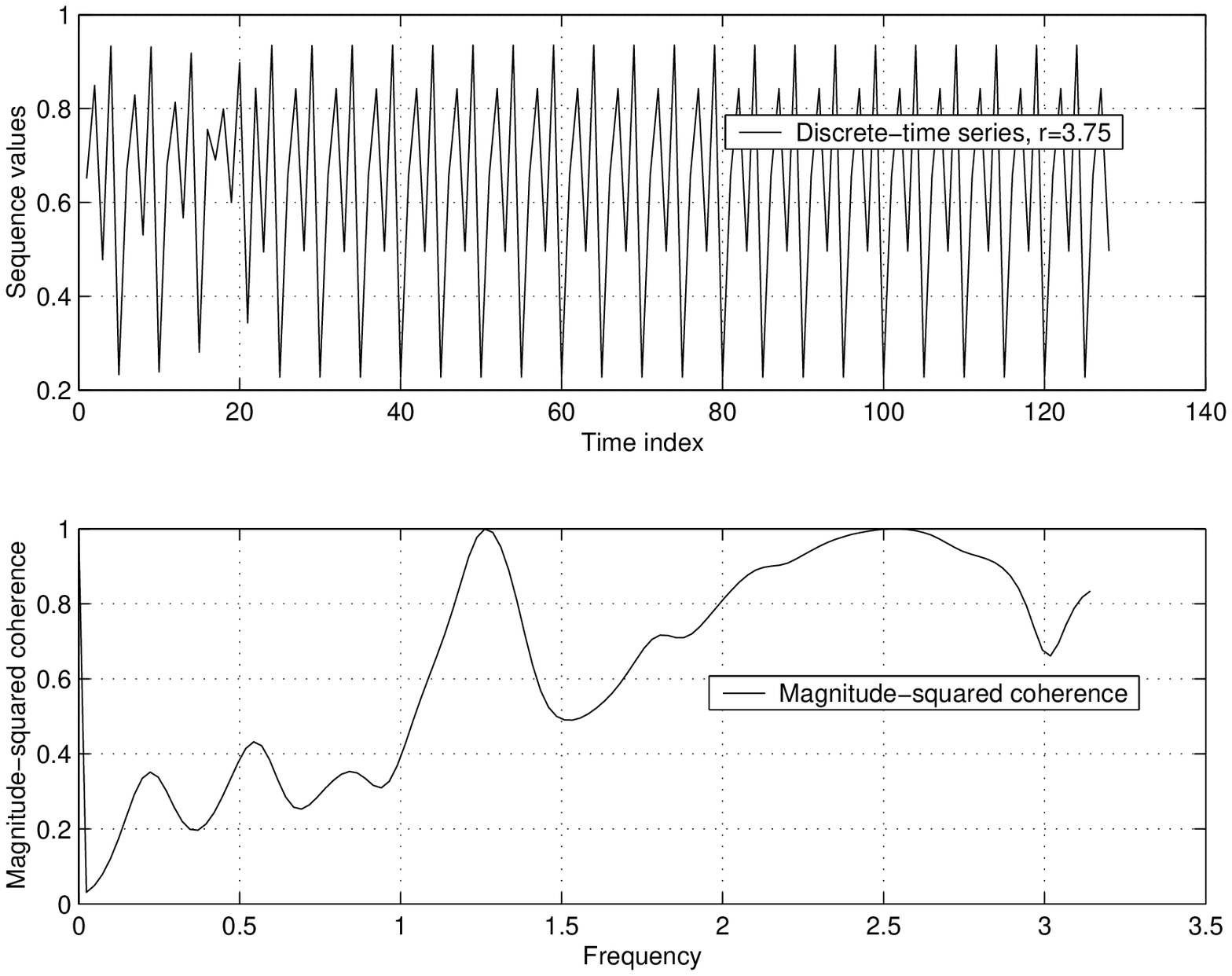,width=0.5\textwidth,angle=0}}
\bigskip
\caption{Discrete-time series and its magnitude-squared coherence,
for r=3.75} \label{fig3}
\end{figure}

\begin{figure}[tbh!]
\centerline{\psfig{figure=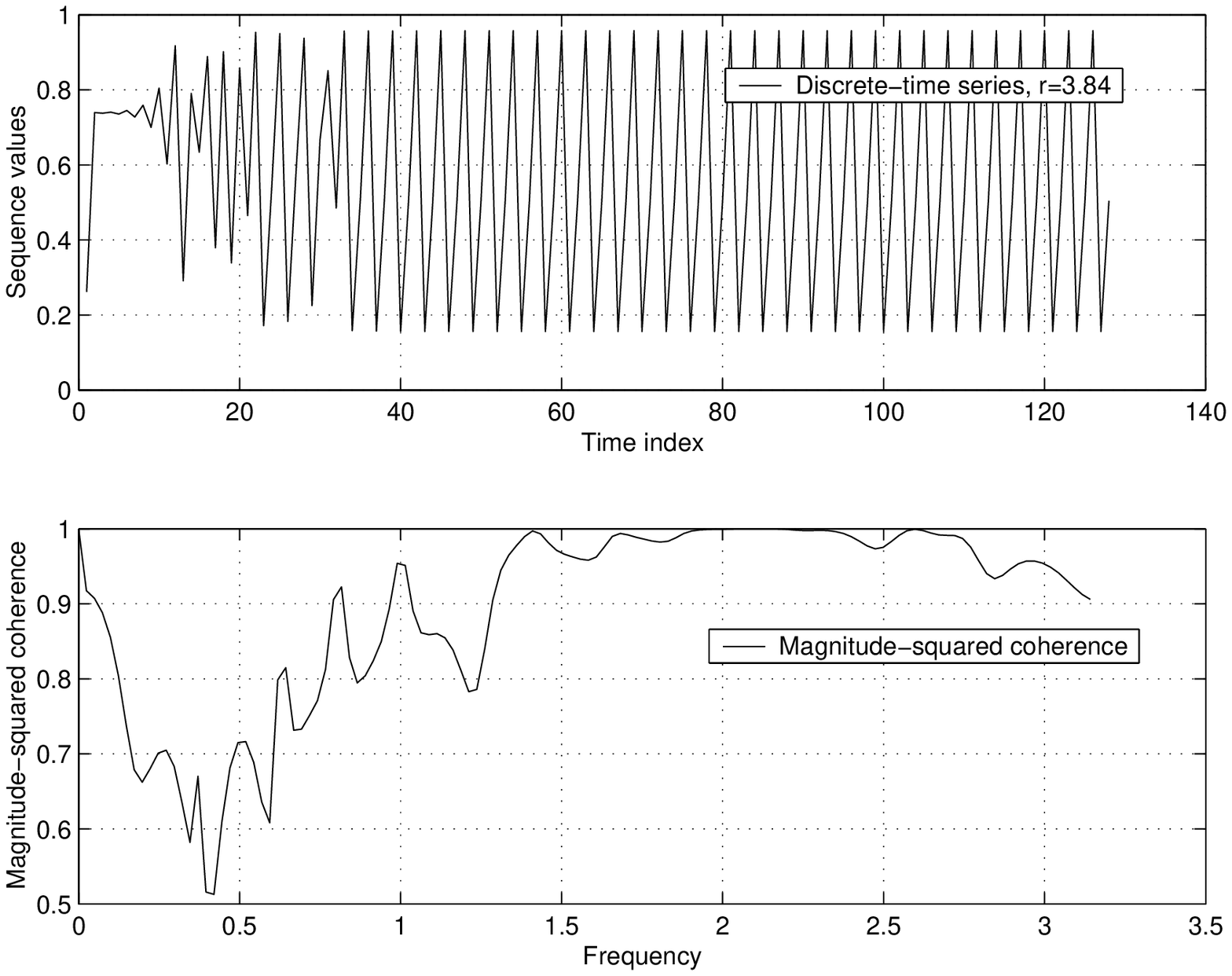,width=0.5\textwidth,angle=0}}
\bigskip
\caption{Discrete-time series and its magnitude-squared coherence,
for r=3.84} \label{fig4}
\end{figure}

\begin{figure}[tbh!]
\centerline{\psfig{figure=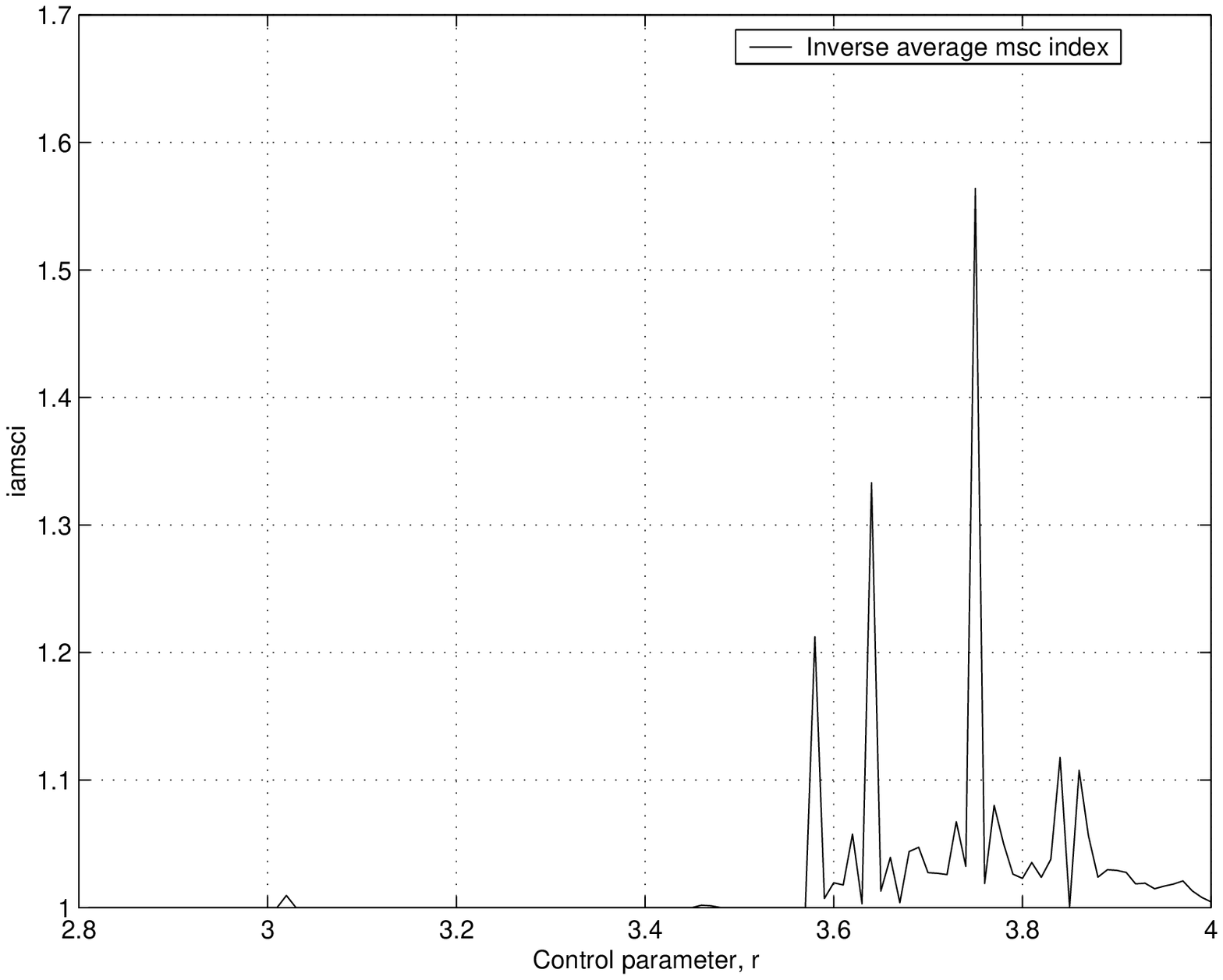,width=0.5\textwidth,angle=0}}
\bigskip
\caption{Inverse average magnitude-squared coherence index}
\label{fig5}
\end{figure}

\begin{figure}[tbh!]
\centerline{\psfig{figure=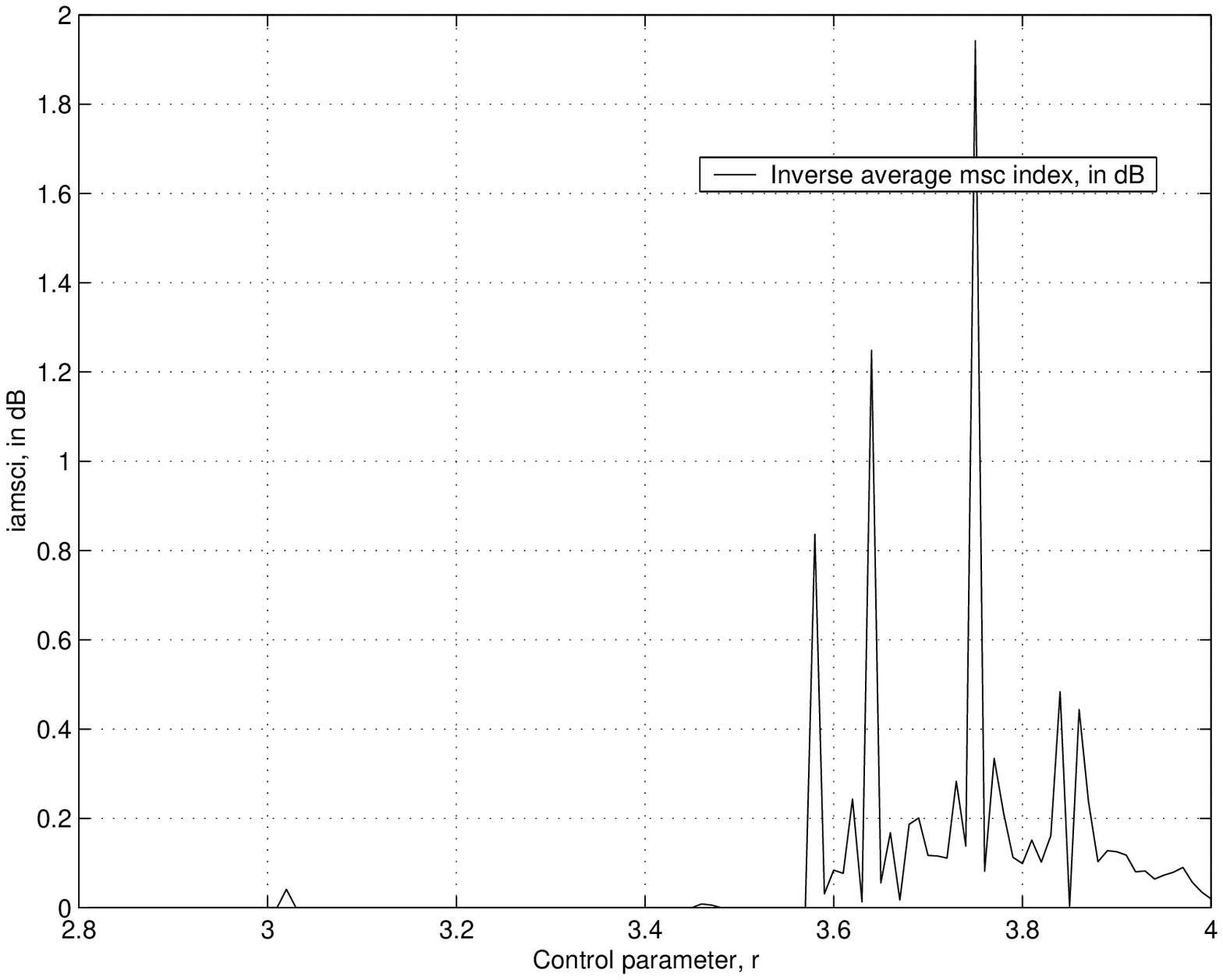,width=0.5\textwidth,angle=0}}
\bigskip
\caption{Inverse average magnitude-squared coherence index, in dB}
\label{fig6}
\end{figure}

\begin{figure}[tbh!]
\centerline{\psfig{figure=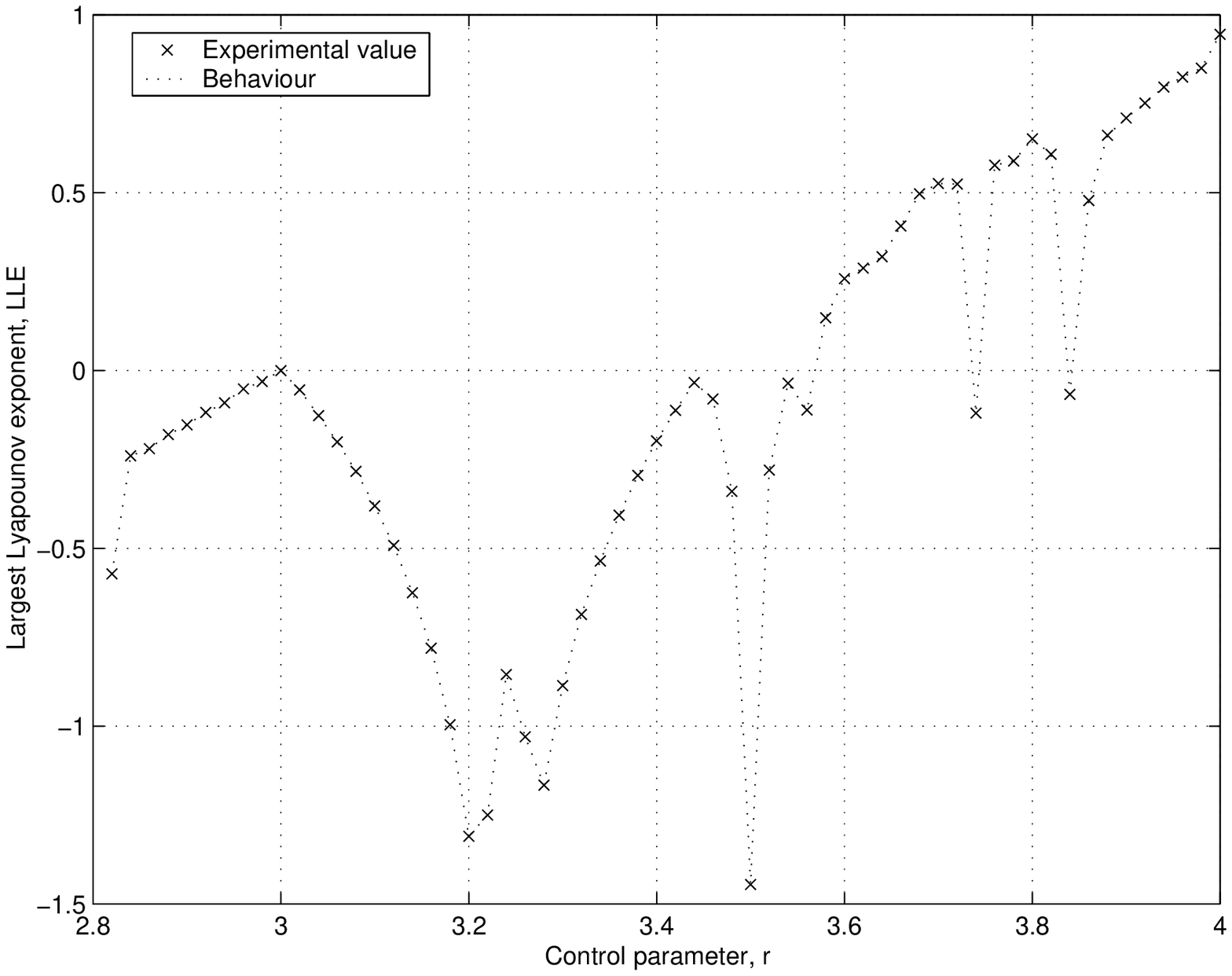,width=0.5\textwidth,angle=0}}
\bigskip
\caption{Largest Lyapounov versus control parameter r}
\label{fig7}
\end{figure}


\begin{references}

\bibitem{Albano}Albano AM, Smilowitz L, Rapp PE, de Guzm\'an GC,
Bashore TR: en Kim YS, Zachary WW (Editors), Physics of Phase Space.
Springer Verlag, Berlin, 1987.

\bibitem{Aradi}Aradi I, Bama G, Erdi P, Groebler T: Chaos and
learning on the Olfactory bulb, Int. J. Int. Sys. Vol.10, 89 117(1995).

\bibitem{Blanco}Blanco S, Figliola A, Kochen S, Rosso O A: Using
nonlinear dynamic metric tools for characterizing brain structures,
IEEE Eng Med Biol, July/August 1997.

\bibitem{Bleek}van den Bleek CM, Schouten JC: Deterministic chaos:
a new tool in fluidized bed design and operation, The Chemical
Engineering Journal, {\bf 53}, 75-78 (1993).

\bibitem{Buzug}Buzug T, Reimers T, Pfister G: Optimal reconstruction
of strange attractors from purely geometrical arguments, Europhys. Lett.13,
605-607,(1990).

\bibitem{Cohen}Cohen ME, Hudson DL, Deedwania PC: Applying Continuous
Chaotic Modeling to Cardiac Signal Analysis. IEEE EMB Magazine, 15 (5),
97-102, 1996. Cohen ME, Hudson DL, Anderson MF, Deedwania PC:
A conjecture to the solution of the continuous logistic
equation. Int J of Uncertainty, Fuzziness and Knowledge-Based Systems 2 (4):
445-461, 1994.

\bibitem{Eckmann}Eckmann JP, Oliffson Komphorst S, Ruelle D,
Ciliberto DS: Lyapounov exponents from times series,
Phys. Rev. A 34 (1986) 4971. Eckmann JP and Ruelle D:
Ergodic theory of chaos and strange attractors, Rev. Mod. Phys. 57(1985) 617.
Eckmann JP and Ruelle D: Fundamental limitations for
estimating dimensions and Lyapounov exponents in dynamical systems. Physica
D 56 (1992) 185-187.

\bibitem{Eddins}Eddins DA, Wright BA: Comodulation masking release
for single and multiple rates of envelope fluctuation, J.Acous.Soc.Am.96(6),
December 1994, 3432-3442.

\bibitem{Fadragas}Fadragas CR, Orozco-Morales R, Lorenzo-Ginori JV:
Possibilities of the discrete Fourier transform for determining
the order-chaos transition in a system governed by the logistic
equation dynamics, nlin.CD/0203010 (Los Alamos DataBase).

\bibitem{Farmer}Farmer JD, Sidorowich JJ: Predicting Chaotic Time Series,
Phs.Rev.Lett.{\bf 59}(8), 845 (1987).

\bibitem{Fojt}Fojt O, Holcik J: Applying nonlinear dynamics to ECG signal
processing, IEEE EMB, 96 (1998).

\bibitem{Fraser}Fraser AM, Swinney HL: Independent Coordinates for Strange
Attrators from Mutual Information. Physical Review A, 33 (1986) 1134-1140.

\bibitem{Grassberger}Grassberger P, Procaccia I: On the characterization
of strange attractors, Phys. Rev. Lett. {\bf 50}, 346 (1983).
Grassberger P and
Procaccia I: Measuring the strangeness of strange attractors, Physica D
9(1983) 189; Grassberger P and Procaccia I: Estimation of the Kolmogorov
entropy from a chaotic signal, Phys. Rev. A28 (1983) 2591.

\bibitem{Hoyer}Hoyer D, Schmidt K, Bauer R, Zwiener U, Koeler M,
Liuthke B, Eiselt M: Nonlinear Analysis of Heart Rate and Respiratory
Dynamics, IEEE
EMB,vol.16 (1), 31 (1997).

\bibitem{Kennel}Kennel MB, Isabelle S: Method to distinguish possible
chaos from colored noise and to determine embedding parameters, Phys.
Rev. {\bf A46} (6), 3111 (1992); Kennel MB, Brawn R, Abarbanel HDI:
Determining embedding dimension for phase-space reconstruction using a geometrical
construction, Phys. Rev. A 45 (1992) 3403.

\bibitem{Liberati}Liberati D, Cursi M, Locatelli T, Comi G, Cerutti S:
Total and partial coherence analysis of spontaneous and evoked EEG by
means of multi-variable autoregresive processing, Med.Biol.Eng.Comp.,
March 1997, 124-130.

\bibitem{Lovett}Lovett EG, Ropella KM, Discrimination of ventricular
arrhythmias from surface ECG via mean magnitude-squared autobicoherence,
IEEE...,1/9,721-722.

\bibitem{Suresh}Narayanaswamy S, Berbari EJ, Lander P, Lazarra R:
Signal processing techniques to determine the mechanisms of premature
ventricular beats, IEEE.....,1/93, 709-710.

\bibitem{Rosenstein}Rosenstein MT, Collins JJ, De Luca CJ: Reconstruction
expansion as a geometry-based framework for choosing proper delay times,
Physica D 73(1994)82]. Rosenstein MT, Collins JJ, De Luca CJ: A practical
method for calculating largest Lyapounov exponents from small data sets.
Physica D 65(1993) 117.

\bibitem{Sih}Sih HJ, Sahakian AV, Arentzen CE, Swirym S: A frequency
domain analysis of spatial organization of epicardial maps, IEEE Transactions
on Biomedical Engineering, Vol.42, No.7, July 1995, 718-727.

\bibitem{Sprott}Sprott JC and Rowlands G, Chaos Data Analyzer, Professional
Version, 1995.

\bibitem{Takens}Takens F: in Lectures Notes in Mathematics, vol.898, Springer,
New York, 366 (1981).

\bibitem{Thakor}Thakor NV, Webster JG, Tomkins WJ: Estimation of QRS complex
power spectra for design of a QRS filter, IEEE Transactions on Biomedical
Engineering, Vol. BME 3, No. 11, November 1984, 702-706.

\bibitem{WolfE}Wolf Emil: Radiometric Model for Propagation of Coherence,
Optics Letters, Vol.19, No.23, December 1, 1994, 2024-2026.

\bibitem{Wolf} Wolf A, Swift JB, Swinney HL, Vastano JA: Determining
Lyapounov Exponents from Time Series, Physica 16D, 285-317 (1985).

\bibitem{Yang}Yang ZA, Chen SG, Wang GR: Determining the delay time of the
two-dimensional reconstruction for ordinary differential equations, Modern
Physics Letters B vol.9, no.19 (August 20, 1995), 1185-1198.
\end{references}
\end{document}